\pdfoutput=1
\documentclass[pageno]{jpaper}

\usepackage[normalem]{ulem}
\usepackage{times}

\usepackage{datetime}
\usepackage{url}
\usepackage{color}
\usepackage{graphicx}
\usepackage{amssymb,amsfonts}
\usepackage{amsmath}
\usepackage{amsthm}
\usepackage{booktabs}
\usepackage{listings}
\usepackage{color}
\usepackage{beramono}

\begin{document}

\title{Automatic Parallelization of Sequential Programs\\
\normalfont\large{Peter Kraft, Amos Waterland, Daniel Y Fu\\
Anitha Gollamudi, Shai Szulanski, Margo Seltzer\\
Harvard John A. Paulson School of Engineering and Applied Sciences}
}

\date{}
\maketitle

\thispagestyle{empty}

\begin{abstract}

\vspace{5mm}
Prior work on Automatically Scalable Computation (ASC)
suggests that it is possible to parallelize sequential computation
by building a model of whole-program execution, using that model to predict
future computations, and then speculatively executing those future
computations. Although that prior work demonstrated scaling,
it did not demonstrate speedup, because it ran entirely
in emulation. We took this as a challenge to construct a hardware
prototype that embodies the ideas of ASC, but works on a broader range
of programs and runs natively on hardware.  The resulting system
is similar in spirit to the original work, but differs in practically
every respect.

We present an implementation of the ASC architecture that runs
natively on x86 hardware and achieves near-linear speedup up to
44-cores (the size of our test platform) for several classes of
programs, such as computational kernels, map-style programs, and
matrix operations. We observe that programs are either completely
predictable, achieving near-perfect predictive accuracy, or totally
unpredictable, and therefore not amenable to scaling via ASC-like
techniques. We also find that in most cases, speedup is limited
only by implementation details: the overhead of our dependency
tracking infrastructure and the manipulation of large state spaces.
We are able to automatically parallelize programs
with linked data structures that are not amenable to other forms
of automatic parallelization.

\end{abstract}

\section{Introduction}
\label{section:intro}

Since the introduction of the first machine with multiple processing units,
automatic parallelization has been the holy grail of scalability.
With the end of Dennard scaling~\cite{dennard74}, enabling software to take
advantage of multiple cores has become even more urgent.
We adopt the Automatically Scalable Computation (ASC) architecture, which transforms
the problem of automatic parallelization to one of prediction and
speculation~\cite{waterland2014asc}.
The original ASC implementation ran on a software emulation of x86; thus,
although it demonstrated promising scaling capabilities, it failed
to achieve true speedup relative to native hardware execution.  Its gains were
therefore ultimately theoretical; the paper did not show that ASC could work outside of simulation.
We present a prototype of ASC, NewAge, that runs on native hardware.  While
NewAge draws inspiration from the original work, its implementation is completely different.
NewAge addresses the major technical challenges left open in the prior work;
addressing these challenges enables true speedup and execution
orders of magnitude faster than that of the original ASC implementation.
NewAge demonstrates the viability of the ASC approach, showing that
it can be used to achieve automatic parallelism and true speedup on
native hardware.
For the remainder of this paper, we use ASC to refer to the general
architecture and NewAge to refer to our current implementation of that
architecture.

ASC achieves automatic parallelism by treating computation as a dynamical
system.
The memory and registers of a program form a (very) large state
vector, and the instruction set architecture acts as a transition rule that
moves the computation from one state to another.
In this model, program execution carves out a path or trajectory through
the state space.
ASC's parallelization strategy is to \textit{predict} points that fall on that
execution trajectory and dispatch \textit{speculative executions} from those
predicted points to available cores.
If the actual computation reaches a predicted state, it can then fast-forward to the
point at which the speculative execution completed.   Crucially, these fast-forwards
happen only if the predictions are correct, so there is no risk of bad execution.
Thus, ASC can speed up program execution by using available cores to
perform some computation early and caching the results until they are useful.

Designing an implementation of ASC requires addressing three fundamental
questions:
1) How do we determine points in a program amenable to prediction and therefore candidates at which to begin speculative execution?
2) What technique(s) should we use to build models of the evolution of computational states?
3) How do we store these speculative executions so that we can look them up efficiently?
We discuss the solutions to all these problems in more detail in Section~\ref{section:implementation} but summarize them here.

We address the first problem by running the target program under a subsystem called the \textit{recognizer} that ranks a set of candidate 
instruction pointer (IP) locations (\S~\ref{section:recognizer}).
After running the recognizer, NewAge uses \texttt{ptrace} and the \texttt{perf\_event} API to set breakpoints
in the target process.
When NewAge regains control at these breakpoints, it captures the process
state for model building and prediction.

We address the second by building two learners for model-building:
a neural network~\cite{lecun-98b} that can be trained online (i.e., we can
both build models and make predictions during the same execution) and an offline-trained collection
of decision trees (these are trained over a few executions before being used to make predictions).
Both attain high accuracy, and each has its advantages and disadvantages
(\S~\ref{section:treesdescription}).

The third problem is perhaps the trickiest aspect of ASC and deserves
further elaboration.
Abstractly, the computational cache is a lookup table from state vectors to
state vectors; given a predicted state, we wish to look up the result of the
speculative computation that started in that state.
A naive cache implementation uses the entire predicted state vector as its
key and stores the entire speculated state vector as the value corresponding to that key.
This is problematic because
these vectors are extraordinarily large.
We can simplify the problem by recognizing that \textit{a cache lookup need not match on all the bits in the state.
It needs to match only  on the bits on which the computation actually depend.}
There is no need to predict or match on bits that do not change or are never read.
We use dynamic instrumentation (using Pin \cite{luk2005Pin})
to identify precisely the set of bits on which a speculative computation depends.
We then use this information to construct cache entries that match any state
that includes correct predictions for this (relatively small) number of
bits (\S~\ref{section:newcache}).

A second challenge in cache design is construction of a unique key to
efficiently identify candidate cache entries.
The original ASC prototype provided no true key mechanism, searching the
cache iteratively on every lookup.
We discovered that \emph{the contents of the live registers\footnote{A register is live at an instruction pointer (IP) if the value of the register at the IP is used before it is overwritten. A register is not live at an IP if its value is always overwritten before the register is used.} provide a
sufficiently unique fingerprint for a program's state that we can use it as
a cache key}.
This allows us to hash the live register contents to produce a key that
with high probability
uniquely identifies a cache entry (followed by a full comparison in the case
of a match).
The combination of dependency analysis and register fingerprinting enables us
to produce an implementation of the ASC architecture that runs natively
on x86/Linux platforms and achieves speedup linear in the number of cores
on several classes of programs.
%
%The contributions of this work are:
%\begin{itemize}
%\item{The NewAge implementation of the ASC architecture that
%produces near-linear speedup on real hardware.}
%\item{An empirical evaluation of NewAge,
%demonstrating which aspects of the system dictate overall speedup,
%what programs are amenable to ASCing, and how this class
%of programs differs from those that can be parallelized via
%compiler-only approaches.}
%\item{A demonstration that ASC is compatible with, and can benefit from, compiler information, by leveraging a \texttt{clang}~\cite{llvm} pass to produce live register analysis.}
%\item{The discovery that the contents of live registers provide a unique
%fingerprint that enables efficient lookup into a computational cache
%~\cite{waterland13caches}.}
%\end{itemize}

Overall, the major contribution of this work is the NewAge implementation of the 
ASC architecture that produces near-linear speedup
on real hardware.  We also provide an empirical evaluation of NewAge,
demonstrating which aspects of the system dictate overall speedup,
what programs are amenable to ASCing, and how this class
of programs differs from those that can be parallelized via
compiler-only approaches.  NewAge is the first implementation of ASC that can run natively
and produce real speedup.  The original ASC could only function in an extremely slow and limited simulation and could only achieve speedup relative to programs run in that simulation.

\section{Background}
\label{section:ascoverview}

To provide context for the discussion of our implementation in the
next section, we briefly review the ASC architecture and terminology.

\begin{figure}
\includegraphics[width=\linewidth]{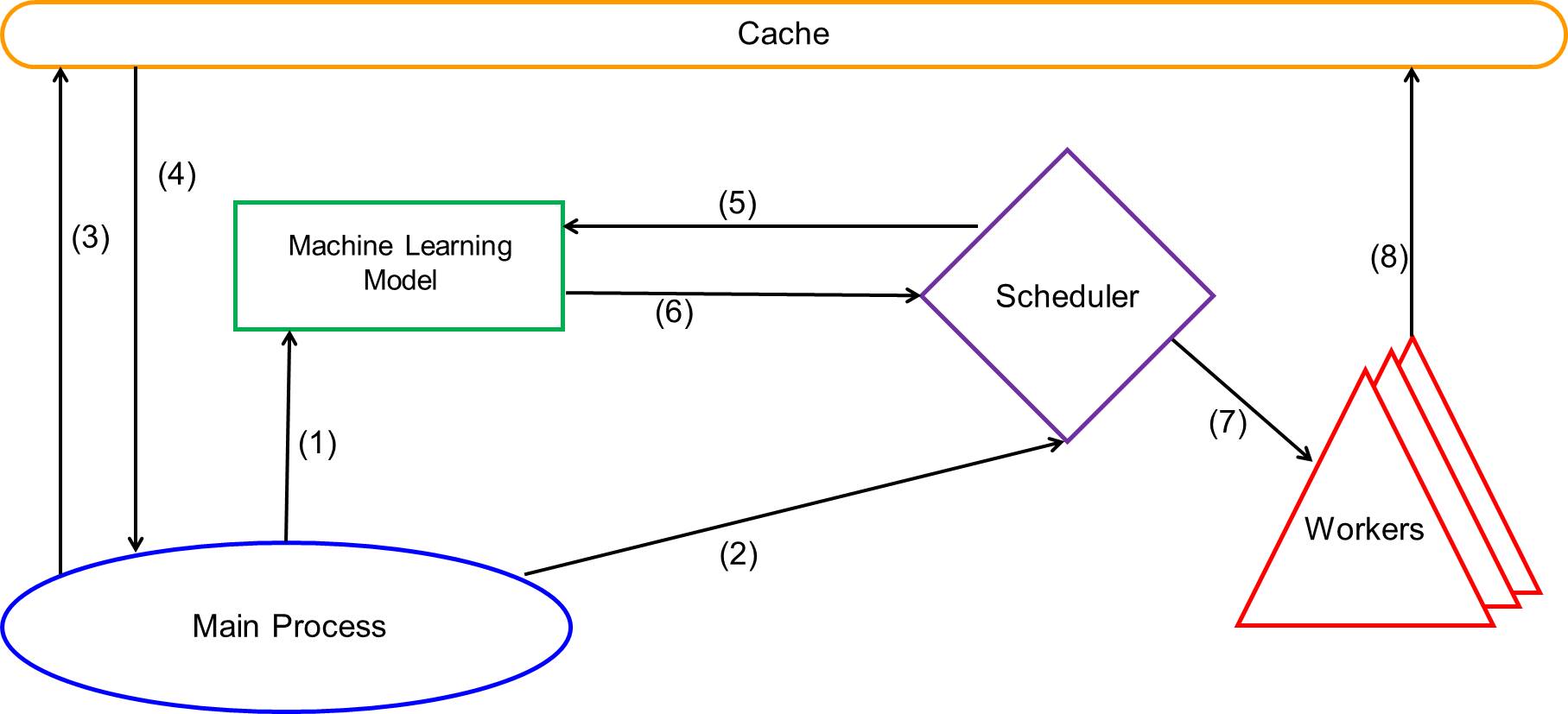}
\caption{\small ASC architecture.
At breakpoints, ASC captures the main process state to train the machine learning model (1)
and provides a basis for prediction to the scheduler (2).
ASC also queries the cache for a match with the current state (3).
On a hit, the cache returns the corresponded speculated state, thereby
fast-forwarding the main process (4).
On a miss, the scheduler uses the current state provided by the main process to request a state prediction (5) and receives back a predicted state (6).
It then dispatches workers to the predicted state(s) for speculation (7).
When the workers complete their speculation, they add an entry to the cache (8).}
\label{fig:architecture}
\end{figure}

We call the program we wish to speed up the \emph{main process}.
Speculations execute in \emph{worker processes}.
The ASC \emph{scheduler} is responsible for requesting predictions
from the machine
learning models and scheduling speculations from those predictions.
We call the states that the models predict the \emph{predicted states}.
When the scheduler gives the workers a prediction, the workers execute a
minimum number of instructions (calculated to produce speculations long
enough to amortize system overhead) and then complete their speculation
at a designated value of the instruction pointer.
The state at which the worker stops execution is called the \emph{speculated state}.
Workers place the pair of predicted and speculated states into the cache as a single
\emph{cache entry}.
Figure~\ref{fig:architecture} (drawn heavily from Waterland et al.~\cite{waterland2014asc})
shows the overall ASC architecture and execution paths.

So far, we have assumed that ASC knows the location in the main process for
which it should build models and speculate.
The process of identifying this location is called \emph{recognition} and
the location at which we build models is called the \emph{recognized instruction pointer} or
\emph{RIP}.
For the rest of the discussion in this section, assume that the ASC user specifies
the RIP; in Section~\ref{section:implementation} we will discuss the subsystem that locates the RIP.

We next describe the components of NewAge in more detail.

\subsection{Machine Learning Model}\label{section:backgroundmlm}

The job of the machine learning model is to make predictions about future states of the main process.
The model takes as input the current state of the main process -- its register
and memory contents -- and returns as output a prediction of the state the
main process will be in the next time it reaches the breakpoint.  To be specific, it predicts
which bits will have changed their values the next time the main process reaches the breakpoint.
At first blush this seems impossible -- the process state is potentially
billions of bits.
However, two things make the problem tractable.
First, many bits in the process state never change, e.g., code.
Second, a worker's computation depends only upon those bits actually
read during its execution.
We can reduce the predicted state further by observing that
any locations written before being read by the worker (i.e.,
scratch space) also need not be predicted.
If we can accurately identify the bits on which a computation
depends, we can significantly reduce the scope of the
prediction problem.
We show the magnitude of this reduction in Section~\ref{section:results}
and describe how we use Pin to accomplish this in Section~\ref{section:implementation}.

\subsection{The Cache}

The cache is indexed by a predicted state and returns a speculated state.
However, as we just discussed, we wish to minimize the size the predicted state,
by including only those parts on which the computation depends.
If each cache entry depends on different bits, how can we efficiently conduct
lookups?
We observe that the contents of the live registers
provide a relatively
unique fingerprint of a process' state and use that to index our cache.
We discuss this in more detail in Section~\ref{section:implementation}.

\subsection{The Scheduler}

The scheduler determines from which (predicted) state
each worker should execute.
Its job is to balance the desire to use all available cores
with the possibility of wasting work.
It must also ensure that workers complete their speculations and
enter them into the cache before the main process encounters their predicted
states.
If workers have not completed their speculation yet, there will be no cache
entry when the main process enters the predicted state and speculation will be wasted.

Finally, the scheduler must also decide how to predict states far in the future.
There are (at least) two ways to do this.
One way is to predict the next state, $S(t)$, feed that prediction back into the learning
model and then predict state $S(t+1)$, etc.
The other way is to ask the machine learners to build multiple models: one that
predicts every state at the RIP, one that predicts every second state, one that
predicts every third state, etc.
Although we have shown that the latter approach is promising,
the results in Section~\ref{section:results} all use the former approach.

\section{Implementation}
\label{section:implementation}

Our ASC implementation, NewAge, is a monitor process that creates the
main process using \texttt{fork} and \texttt{exec} and all workers 
(typically one for each available core) using Pin. 
NewAge uses \texttt{ptrace} and the \texttt{perf\_event} API on the main process and Pin instrumentation on the workers 
to set breakpoints at the RIP.
At this point, the main process begins running.

Every time the main process reaches the breakpoint,
NewAge looks up the current state in the cache, fast-forwarding on a hit.
NewAge also presents the current state to the machine learning model as input and
obtains a prediction for a future instance of the RIP.
If there are workers available, NewAge will then dispatch a worker
to speculate from the newly predicted state.
If NewAge constructs multiple models, it can dispatch multiple workers, one
at each predicted state.
This cycle repeats until the main process completes execution.

In the following sections, we discuss the major components of NewAge.

\subsection{The Recognizer}
\label{section:recognizer}
We currently perform recognition by running the target program twice - once
to identify the RIP, once to identify the period, the number of times we should skip over the RIP before breaking to
create an execution sufficiently long that it amortizes our fixed overheads.
We can then begin building models and/or speculating on subsequent runs of
the program.
On one hand, running the program beforehand feels a bit like cheating; on the other hand, consider
a scenario where we run an expensive computational kernel on a
variety of inputs -- we pay two sequential runs as overhead to locate
optimal breakpoint settings and then speed up all subsequent runs.

NewAge finds the RIP using Pin and a custom Pintool (a shared library called by Pin) that measures the frequency of various
\emph{candidate} instructions, such as jumps and branches (which frequently indicate loops).
We then select the recurring candidate with the lowest frequency above some threshold.
Subsequent runs of the target program instrument
only this breakpoint to collect timing information, selecting a
breakpoint period that balances NewAge overhead with the cost of
execution between two instances of the RIP.
We developed these heuristics after manually
optimizing breakpoints for our test kernels.
NewAge regularly finds the RIP in the same basic block that we
identified manually.

An earlier implementation of the recognizer applied the technique
described by Swersky \emph{et al.}~\cite{adams2014freezethaw}.
We extracted candidate breakpoints from the binary and
trained full instances of ASC incrementally on each breakpoint with
various settings of the parameters, such as minimum instruction count (the minimum
number of instructions we require between breaks).
We then used a Gaussian process regressor to predict the speedup
possible with additional training;
the more promising a setting looked, the longer we trained.
While this implementation successfully identified
good breakpoints, it was expensive to gather training
sets for each candidate.

\subsection{Machine Learning Model}
\label{section:treesdescription}

We have developed two different learners for NewAge.
Both take as input the full state of a program and predict as output the values
 of the bits on which computation depends (identified using Pin) after one timestep, as explained in Section \ref{section:backgroundmlm}.  

The first learner uses a fully-connected 3-layer neural net.
The major advantage of using neural networks is that they can be trained
online, so we can build the model at the same time that we parallelize
(and speed up) the program.
However, they have several disadvantages.
First, they are opaque; we cannot reason about what causes
the networks to produce their output.
Second, they are inflexible in the size of their inputs and outputs.
As we do not know all the bits on which a computation depends a priori, each time we
learn that we want a new bit in the input or output, we have to build a new model.
Third, the time complexity of prediction and training scale quadratically with
the number of bits predicted, because the network is fully connected.
This makes the network too slow to produce speedup for large numbers of bits,
although we can produce speedup for small numbers of bits.

To mitigate these issues, we turn to decision trees trained via the Classification and Regression Tree (CART) algorithm.  These have the downside of needing
offline training, where a set of training examples is gathered from different runs of a program on random inputs and the trees
are batch-trained from it.
However, they map well to this domain, because the training set
consists entirely of binary values (bits) with no noise.
This avoids the two largest disadvantages of decision trees: their
difficulty in splitting many-valued input and their tendency
to overfit to noise.
Furthermore, they offer several advantages.
First, we can train a separate tree for each output bit, avoiding the problem
of rebuilding models when we discover new
bits.
Second, the time complexity of prediction and training increases
linearly, not quadratically, with the number of bits predicted
(as our decision trees have a set maximum depth and therefore
effectively constant size).
Third, trees are transparent, allowing us to reason about
a tree's behavior simply by examining it.

\subsection{Cache}
\label{section:newcache}

The cache must map predicted states to speculated states and provide
efficient lookup.  NewAge has a single shared cache used by the main process and all workers and
protected by locks, unlike the original ASC where each core maintained its own cache requiring every cache query
to produce parallel cache lookups on all cores.
A naive cache simply maps predicted states, $z_p$ to speculated
states $z_s$.
However, we can expand the set of programs on which NewAge can
achieve speedup if we reduce the number of bits it must predict
correctly.
Thus, we wish to augment our cache with masks that
identify every bit on which a speculative execution depends (the
\emph{read-mask}) and every bit written by the execution (the \emph{write-mask}).
These masks play the same role as those introduced by
Waterland et al~\cite{waterland2014asc}; we review their use
here for completeness.

Our cache implements two functions, \texttt{cache\_add} and
\texttt{cache\_lookup}.
Function \texttt{cache\_add} simply takes the four-tuple $(z_p, z_s, m_r, m_w)$,
where $m_r$ and $m_w$ are the read- and write- masks respectively,
and adds it to the cache.
Function \texttt{cache\_lookup} takes the main process state, $z_m$,
and checks if any cache entry contains a predicted state $z_p$ identical
to $z_m$ \emph{in all bits read during the computation of $z_s$}; that is,
if: $z_m \wedge m_r =  z_p \wedge m_r$.
If an entry matches, the cache returns both $z_s$ and $m_w$, else
it returns NULL.

\begin{figure}
\centering{\includegraphics[width=0.85\linewidth]{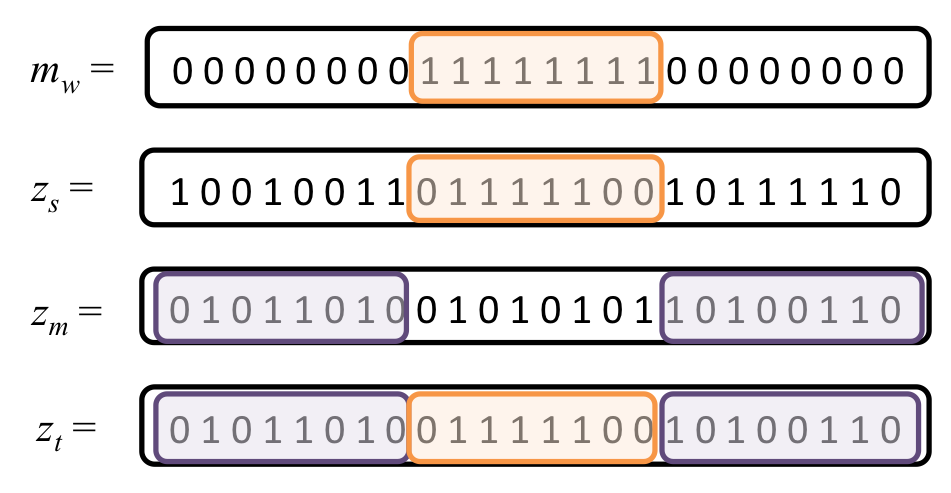}}
\caption{\small Constructing a fast-forwarded state from a cache entry.
NewAge has found a cache entry, containing $z_s$ (the speculated
state) and $m_w$ (the write-mask),
that matches the main process state $z_m$ under the read-mask $m_r$ (not displayed).
It constructs future state $z_t$ by copying only the bits under the write-mask
in $z_s$ (orange) into $z_m$, leaving the other bits untouched (purple).
The final construction $z_t$ contains the first and third bytes of $z_m$ and
the second byte of $z_s$.}
\label{fig:boolconstruct}
\end{figure}

On a cache hit, we must construct the state, $z_t$ that results from
fast-forwarding $z_m$ using the cache entry.
Logically, we want to take the bits in $z_s$ under the mask $m_w$
and insert them into $z_m$ to produce $z_t$.
Algebraically, we want:
$z_t = (z_s \wedge m_w) \vee ((z_m \vee m_w) \veebar m_w)$, where $\veebar$ denotes the
bitwise XOR.
The first clause produces the bits written during the speculation,
and the second clause produces a version of the current state
with every bit under the write-mask set to 0.
Figure~\ref{fig:boolconstruct} illustrates this lookup process.

We implement the cache as a hash table.
However, doing so requires a fast hash function that does not
depend on $m_r$, since every entry in the cache could
conceivably have a different $m_r$.
After much experimentation, we have discovered that the values of
the live registers are sufficiently unique to act as a fingerprint
for a process's state at a RIP.
While the set of memory values that are read from might change
between iterations of a program, the set of registers that are
live at the breakpoint never changes.
It is part of every $m_r$.
Moreover, the values of the registers do a good job of
discriminating states:  for all programs we tested,
if the states of the live registers were identical,
the states of the areas of memory under the read-mask were
also identical.
Given the centrality of registers to all computations,
this is likely true in general for non-adversarial programs.
This has allowed us to implement the cache as a hash table where the ``hash''
of each tuple $(z_p, z_s, m_r, m_w)$ is a simple hash function
over the values of the live registers of $z_p$.
When we lookup a main process state $z_m$, we simply hash its
live registers and look up the hash in the table.
This is equivalent to asking if there
is an entry $(z_p, z_s, m_r, m_w)$ where the register values of
$z_p$ and $z_m$ are identical.  We do not rely on this approximate hash, however.
If we do find such an entry $(z_p, z_s, m_r, m_w)$ we verify
that $z_m \wedge m_r =  z_p \wedge m_r$, which
is equivalent to making sure that the full state vectors $z_m$ and $z_p$ are identical
under the read-mask.  If the equality is true, we have a hit.  That last equality check makes this scheme
resilient to any possibility of hash collisions (where the live register set is the same, but memory values differ).  They may lead to
the loss of a cache entry and some wasted computation, but never to invalid outputs or aberrant behavior.

The astute reader may recognize that it is sometimes necessary
to perform multiple sequential cache lookups.
This occurs when the final modified state retrieved from the cache
matches some other cache entry.
Depending on the scheduler, this can happen many
times in succession.
Originally, we solved this problem by having \texttt{cache\_lookup} iterate.
If it found a speculated state, instead of returning it would look up that
speculated state as a predicted state and repeat until it had reached a
speculated state that had no match in the cache.
However, this solution becomes
extremely expensive when state vectors became large, since it requires
synchronous performance of a large number of whole-state-vector
Boolean operations in the main process.
We remove this iterated lookup by having asynchronous threads
optimize the cache.
We search for pairs of entries $(z_p, z_s, m_r, m_w)$ and
$(z_p', z_s', m_r', m_w')$ such that $z_s \wedge m_r' = z_p' \wedge
m_r'$; that is, pairs of entries where the speculated state of the
first matches the predicted state of the second under the read-mask
of the second.
If we find such a pair, we combine it into a single
entry $(z_p, z_t, m_r \vee m_r', m_w \vee m_w')$ where $z_t$ is the
state constructed from $z_p$ and $z_s'$ using the procedure described
above and combining the read- and write- masks.
We call this optimization \emph{cache stitching}.

\paragraph{Live Register Analysis} During the execution of the program,
we must know the set of registers that are live at a program point
in order to predict the future state.
However, \texttt{clang} erases register liveness information right after the
register allocation pass. We have therefore modified the \texttt{clang} toolchain
to compute the liveness information after the physical registers allocation pass but before
the emittance of assembly instructions.
We use the standard \emph{backward propagation} technique to compute the live range
of registers.
Since our kernels are non-adversarial,
we assume that the instructions in the linked binary
are in one-to-one correspondence with the instructions emitted during the assembly phase, enforcing
that assumption by detecting instruction padding for alignment and disabling link time optimizations for kernels.

\subsection{Scheduler}
\label{section:newscheduler}

The goal of the scheduler is to assign different predictions to
different workers so that every worker is usefully speculating about
the future of the main process.
We define a \emph{timestep}, $t_i$ as the $i^{th}$ instance that
NewAge stops the main thread at the RIP.
The first time NewAge stops the main thread at RIP is $t_1$,
the second is $t_2$, and so on.
We use timesteps to simplify the scheduler by recognizing that
the predictions created by our machine learning model are not
only general predictions for the future but also predictions
for a specific timestep.
The scheduler maintains a large table with an entry for each timestep
(modulo a large number).
If a worker has completed speculation or is currently speculating
a particular timestep, it marks the appropriate entry in the table.
Every time a worker process is ready for
scheduling, the scheduler assigns it to the $N^{th}$ unmarked entry
in the table, where $N$ is a small number chosen to ensure the
speculation completes before the main thread reaches the timestep.
Assuming the main process was last observed in timestep $t$ and the
chosen entry is for timestep $t'$, the scheduler then has the machine
learning model iteratively predict from the main process state
$t' - t$ times to create a prediction for timestep $t'$.
Next, the scheduler constructs the predicted state and dispatches the worker
to speculate from it.
This ensures that no two workers ever work on
the same prediction and that every speculation from a predicted
state at some timestep $t$ finishes before the main process reaches
timestep $t$.

\subsection{Pintool}
\label{section:pintool}

Our system launches all workers using Pin and a custom Pintool (a shared library called by Pin).
The Pintool contains only a single data structure, a large segment
of memory that is shared with NewAge, Pin's parent.
This segment is large enough to contain copies of the registers, active
areas of the program's memory, and a read-mask and write-mask each the same
size as those active areas.
We determine the active areas of memory and their size by parsing the
main process's memory map in \texttt{/proc/pid/maps}.
The Pintool also acts as a library that contains several
instrumentation functions.
Specifically, we instrument:

\textbf{Memory reads} All instructions that read from memory,
inserting multiple calls if the instruction reads from multiple locations.
The instrumentation function for memory reads takes as arguments the address
read from and the size of the read (in bytes). It then updates
the read-mask.

\textbf{Memory writes} All instructions that write to memory,
inserting multiple calls if the instruction writes to multiple locations.
The instrumentation function for memory writes takes as an argument the
address that was written to along with the size of the write (in bytes). It then updates
the write-mask.

\textbf{The RIP} The instruction at the location of the RIP, where we run
speculative executions and fast-forward the main program.
The instrumentation function takes as an argument a context data structure
containing the values of all the worker process's registers (including flag
registers) at the time the RIP is reached.
The instrumentation function copies those register values as well as the contents
of the program's active memory areas into the shared memory segment, signals NewAge,
and then sleeps.
While the instrumentation function sleeps, NewAge replaces the contents of the shared memory
segment with new memory and register values and then signals the Pintool.
The instrumentation function wakes up and copies the new memory values from the shared memory section into
their proper locations, clears the read- and write- masks, and restarts execution
using the new register and memory values.

\subsection{Scattering and Gathering}

Creating predicted states and recording cache entries requires that NewAge be
able to construct copies of program state.
We accomplish this using \texttt{ptrace} and shared memory to perform scattering and gathering operations on a
process's memory image.

We create a copy of a program's state via a gather operation.  We gather from the main process
by using \texttt{ptrace} to copy in its registers, then using direct memory access (DMA) 
via Linux's \texttt{process\_vm\_readv} to copy in writable segments
of its memory space.  We gather from workers by having them copy the contents of their registers and writable
memory segments into shared memory whenever they reach the breakpoint, then sleep and await scattering.
Scatter is the opposite.  For the main process, we use \texttt{ptrace} to place the
registers that are in the state vector into a process and use DMA via Linux's \texttt{process\_vm\_writev}
to copy the writable memory segments from the vector into the process.  For the workers, we have ASC place
the state vector it wishes to scatter into shared memory, then wake the worker and have it copy the contents of shared memory
into its registers and memory.

\section{Experiments}

\lstdefinestyle{customc}{
  belowcaptionskip=1\baselineskip,
  breaklines=true,
  xleftmargin=\parindent,
  language=C,
  showstringspaces=false,
  basicstyle=\footnotesize\ttfamily,
  keywordstyle=\bfseries\color{red},
  commentstyle=\itshape\color{purple},
  identifierstyle=\color{blue},
  stringstyle=\color{orange},
}

\label{section:results}

The goal of our evaluation is twofold: to determine how effective our
implementation is at achieving speedup and to identify the limitations in
ASC's ability to provide such speedup.
We begin by introducing a set of kernel benchmarks and the methodology we use
to evaluate them.
Next, we demonstrate NewAge's speedup on the collection of kernels.
We deconstruct those results by examining how well
data dependency tracking (the reduction of the state space from all of
memory to the set of bits that change during computation and whose changes affect the results of speculations) reduces state size,
quantifying how much overhead Pin introduces, which we use
to derive the CPU efficiency we achieve in the presence of Pin.
This, in turn, limits the speedup possible by NewAge, and we
examine how close it comes to achieving that maximum speedup.
We report on the efficacy of our predictions and conclude the section
with a discussion of the inherent limitations in ASC.

\subsection{Kernels}

Our benchmark collection draws on three sources: kernels used in the
prior ASC paper, kernels we wrote ourselves, and kernels from
the Polyhedral Benchmark suite (PolyBench/C)~\cite{polybench}.
We briefly describe the prior ASC benchmarks and the benchmarks we
wrote; PolyBench/C is described in detail online~\cite{polybench}.

The kernels adopted from the original ASC~\cite{waterland2014asc} work are:

\textbf{collatz}: A kernel that iterates through a range of
positive integers, testing if each satisfies the Collatz conjecture.
The \emph{Collatz conjecture} states that if one starts with some positive
integer $n$ and sequentially divides it by two if it is even and
multiplies by three and adds one if it is odd, the sequence will
eventually converge to one.
%% MIS: I think we can omit this
%% NewAge parallelizes it by having
%% different workers test the Collatz conjecture on different numbers.
%% This kernel is particularly interesting because NewAge automatically
%% memoizes the results for each integer in a sequence, reusing the same
%% speculation multiple times and thus outperforming conventional
%% parallelization techniques \cite{waterland2014asc}.
%% Note that this automatic memoization is for an arbitrary computation
%% embedded in the program, not a function call, as we typically assume
%% for memoization.

\textbf{ising}: A pointer-based condensed matter physics
program. It iterates through a linked list of spin configurations,
identifying the element in the list with the lowest energy state.
It is interesting because existing parallelizing compilers cannot
parallelize it due to the pointer dereferencing \cite{waterland2014asc}.
NewAge parallelizes it by predicting values referencing later nodes of
the linked list.

%\textbf{mm}:  Naive integer matrix multiplication. It is
%adapted from the \texttt{2mm} multiple matrix multiply kernel in
%Polybench/C~\cite{polybench}.
%NewAge parallelizes it by having different workers compute different
%sections of the final matrix.
%% Unlike the previous kernels, it is extremely memory-intensive,
%% with memory use scaling quadratically with input.
%% It therefore tests overhead on high-memory computations
%% (we used 4000x4000 matrices, meaning each worker process used over a
%% gigabyte of memory).

Next, we run the PolyBench/C~\cite{polybench} kernels, each adapted
slightly to deal with artifacts of the ASC implementation while still performing the same computations.
%% Polybench/C covariance matrix calculation.
%% NewAge parallelizes it
%% by having workers compute different sections of the final matrix.
%% Like \texttt{gemm}, it is extremely memory-intensive.
%% It has nonuniform distance between breakpoints (the breakpoint is
%% reached more frequently over time) because the final covariance
%% matrix is symmetric, so only its upper triangle needs be calculated.

The two new kernels we introduced are:

\textbf{3sum}: A naive cubic solver for the 3-subset sum
problem. It generates a large random list of integers and iterates
through it looking for a set of three integers that sum to zero.
NewAge parallelizes it by having workers test different sets
of integers.
%% Like \texttt{cov}, it has nonuniform distance between breakpoints (the
%% breakpoint is reached more frequently over time) because the size
%% of the search space decreases over time.

\textbf{readmap}: A simple, but computationally
intensive, map operation that repeatedly adds random values to entries
in an array.
NewAge parallelizes it by having different workers perform
the map on different elements of the array.
%% This program demonstrates
%% an embarrassingly parallel problem that ASC should
%% be able to speed up, but cannot without data dependency
%% tracking.

\subsection{Methodology}
\label{section:methodology}

We ran all experiments on a Microway 1U Xeon server with two Intel
Xeon E5-v4 processors. Each of the two processors has 22 cores,
for a total of 44. We disabled Intel Turbo Boost on all cores to
guarantee consistent clock speeds between cores. The server has
256 GB of main memory.
We designed all kernels to randomize behavior
between runs (for example, \texttt{3sum} randomly generates its search space
every run). This randomization demonstrates that NewAge is
actually making predictions about states and not simply memorizing
end states.
The exception to the randomization is \texttt{collatz}, whose nature is not
amenable to randomization. We experimented on \texttt{collatz} by
training it on one set of numbers and testing it on a different set
of numbers.
We used our recognizer (\S~\ref{section:recognizer}) to identify the RIP (the location for which we will make predictions).
We compiled all kernels, except \texttt{readmap}, with \texttt{clang}
using O1 optimization
and the live register analysis described earlier.
We compiled \texttt{readmap} with optimization level O0, because any
dead code elimination pass trivializes it.
We trained all decision trees offline by running each kernel ten times
on different inputs.
We used these trees in all subsequent experiments.  We validate all outputs against the same
programs run natively.

\subsection{Speedup and Analysis} \label{section:scalability_results}

\begin{figure}
\includegraphics[width=\linewidth]{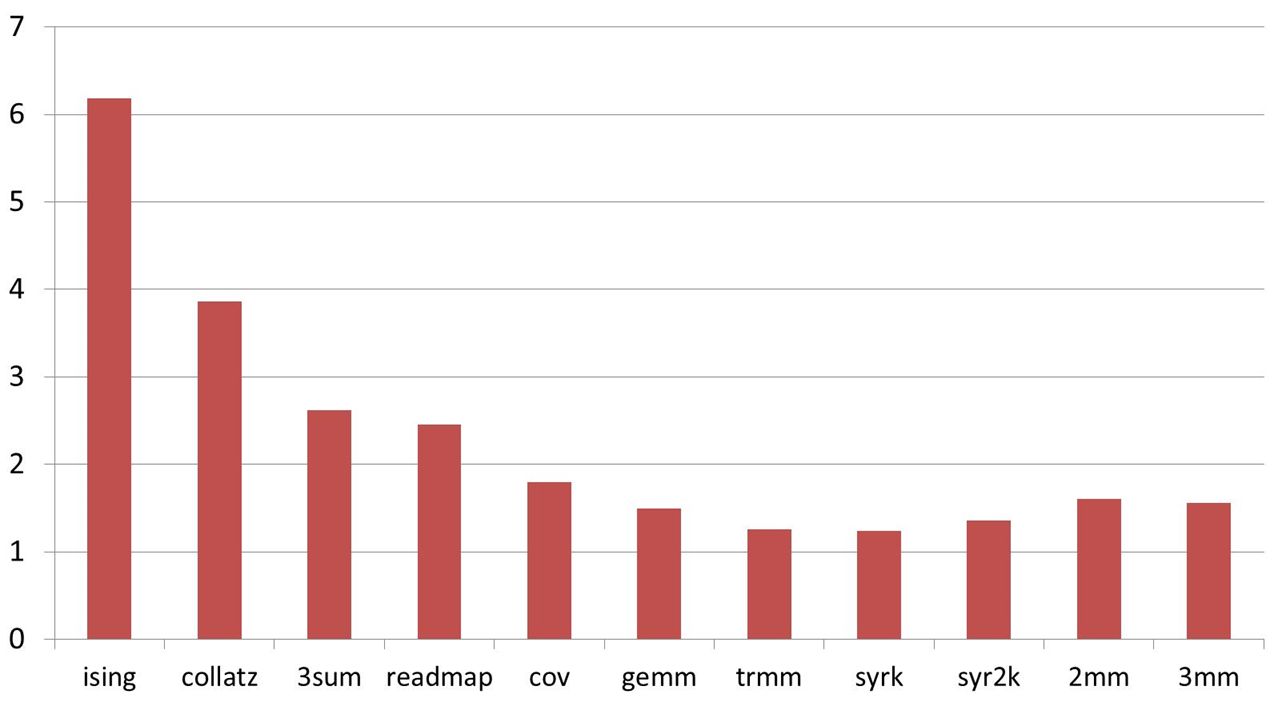}
\caption{\small Speedups attained by all kernels with ten workers.  Speedup is defined as the single-thread uninstrumented execution time
divided by the time it takes running under NewAge with a given number of
workers.}
 \label{fig:speedups_at_10_all}
\end{figure}

Figure \ref{fig:speedups_at_10_all} shows the speedup results for all kernels, both our own and PolyBench/C, for 
which we could attain speedup.  We were able to attain speedup for all of our own kernels plus seven PolyBench/C kernels--\texttt{cov}, \texttt{gemm}, \texttt{trmm}, \texttt{syrk}, \texttt{syr2k}, \texttt{2mm}, and \texttt{3mm}.
The benchmarks for which we were unable to achieve speedup fall into two
categories: those for which our learning model was unable to make accurate
predictions (e.g., \texttt{warshall}, \texttt{cholesky}) and those whose
runtime is small relative to memory size (e.g., 
\texttt{gesummv}, many stencils).
For the latter class, we could make accurate predictions but our overheads due
to memory consumption were too great.
We discuss this in more detail in ~\S~\ref{section:cpuefficiency}.

\begin{figure}
\includegraphics[width=\linewidth]{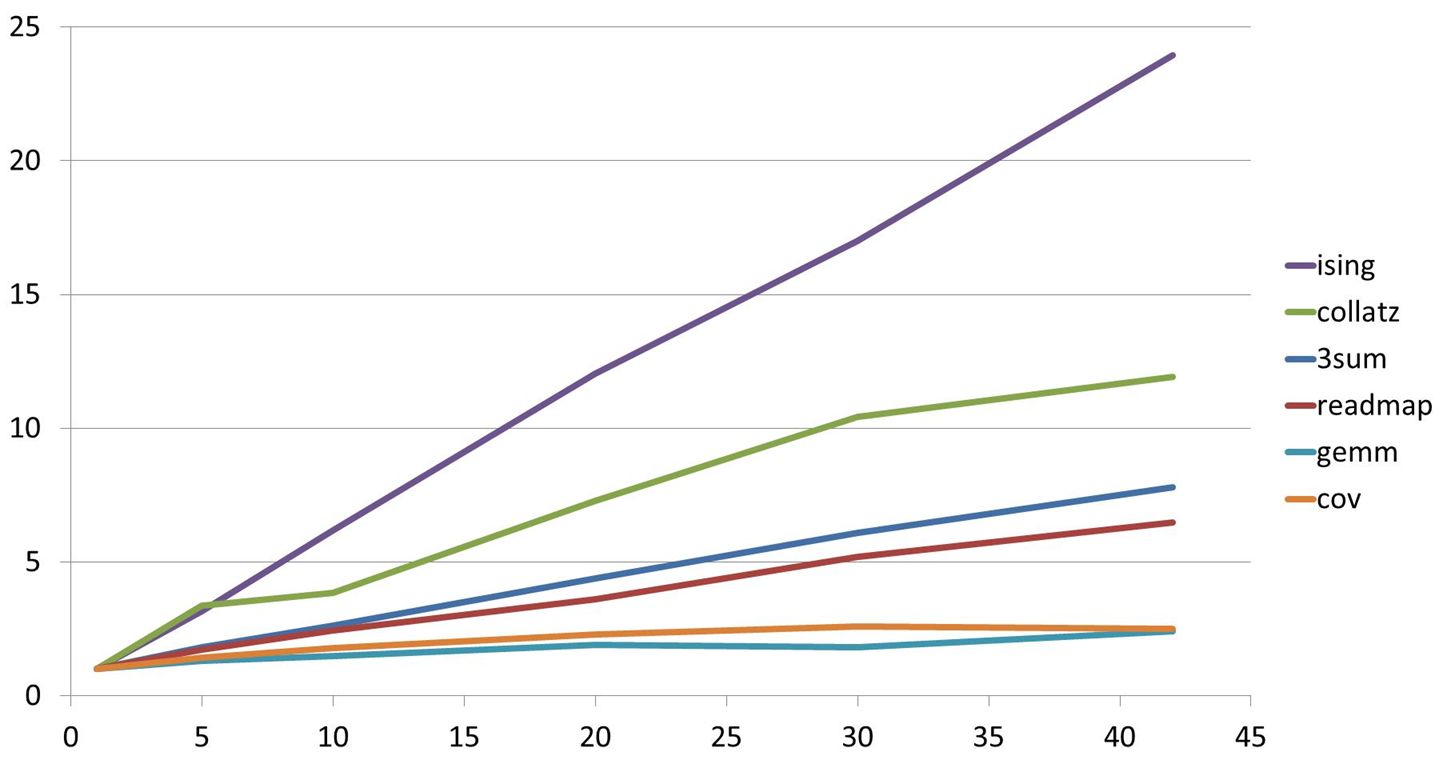}
\caption{\small Speedups attained by selected kernels as a function of the number
of workers. Speedup is defined as the single-thread uninstrumented execution time
divided by the time it takes running under NewAge with a given number of
workers.}
 \label{fig:speedups_all}
\end{figure}

In the interest of conserving space, we restrict our detailed analyses to the performance of six of the kernels each of which demonstrate different aspects of ASC. As we mention below, some of these, e.g., \texttt{gemm}, are representative of a large class of programs.
We defer discussion of the benchmarks for which we are unable to attain speedup
to~\S~\ref{section:asclimits}, where we discuss the limitations of ASC and NewAge.
Figure \ref{fig:speedups_all} shows the speedup results for our six selected kernels,
using 5, 10, 20, 30, and the full complement of 42 workers (reserving one core
for the main process and another for NewAge itself).
Observe that there is a wide range in the scalability achieved for the
different benchmarks with \texttt{ising} using over 50\% of all 44 cores,
while \texttt{cov} achieves far less.
The following sections walk through the factors that determine the
speedup possible, given the NewAge implementation.

\subsection{State Space Size}
\label{section:statespacesizereduction}

\begin{table}
\centering
\begin{tabular}{l r r}
 \toprule
Kernel & Bits w/out DT  & Bits with DT\\
\midrule
\texttt{3sum} & 80 & 80 \\
\texttt{readmap} & 2520 & 120 \\
\texttt{collatz} & 224 & 224 \\
\texttt{ising} & 280 & 280 \\
\texttt{gemm} & 1024000088 & 88 \\
\texttt{cov} & 576000352 & 352 \\
 \bottomrule
\end{tabular}
\caption{Sizes of kernel state spaces without data dependency tracking (DT) and
with data dependency tracking.}
\label{table:statespacereduce}
\end{table}

We begin our performance analysis by examining the state sizes
for each kernel.
We define the size of the
state space induced by our
Recognized Instruction Pointer (RIP)
as the total number of bits whose
values NewAge must predict.  As discussed in Section \ref{section:newcache}, these
are the bits that change
during computation between two occurrences of the RIP
and whose changes affect the results of future speculations.
Table \ref{table:statespacereduce} shows the state space sizes for each
kernel with and without data dependency tracking.
The results reveal two important points.
First, some kernels do not require data dependency tracking
(i.e., \texttt{collatz}, \texttt{ising}, and \texttt{3sum}),
because they do not modify large areas of memory.
Second, while data dependency tracking reduces the state size of \texttt{readmap}
only modestly,
it makes a profound difference for \texttt{gemm} and \texttt{cov}.
Note that the two programs with the large state spaces are also the two
kernels obtaining the least speedup; we'll examine this further in
the next section.

\subsection{CPU Efficiency}
\label{section:cpuefficiency}

\begin{figure}
\centering{\includegraphics[width=0.85\linewidth]{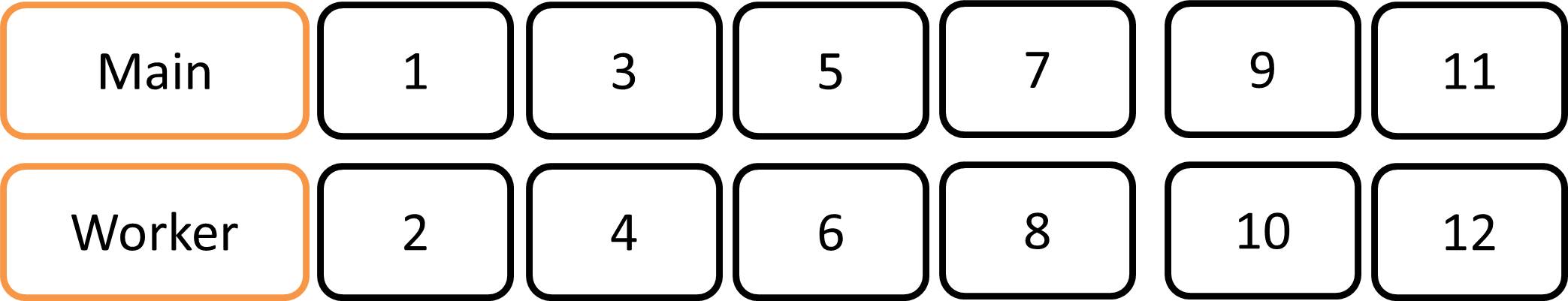}}
\caption{\small ASC with a single worker having 100\% efficiency.
The numbers in the boxes are timesteps.
The main process computes the first timestep while the worker computes
the second.
When the main process completes the first timestep, it fast-forwards through
the second timestep to the third.
While the main process computes the third timestep, the worker computes
the fourth.
Once again, the main process can fast-forward through the fourth timestep
to the fifth.
In this scenario, the main process finishes the computation twice as
quickly as it would natively, achieving 2x speedup with a
50\% computational cache hit rate.}
\label{fig:cpu_efficiency_100}
\end{figure}

Next, we examine Pin's overhead and the resulting computational efficiency
we obtain from our workers.
Let's begin by assuming that we have an ideal model that always produces
accurate predictions and that Pin introduces no overhead.
In this scenario, depicted in Figure~\ref{fig:cpu_efficiency_100},
workers run at exactly the same speed as the main process; we
call this 100\% CPU efficiency.

\begin{figure}
\centering{\includegraphics[width=0.85\linewidth]{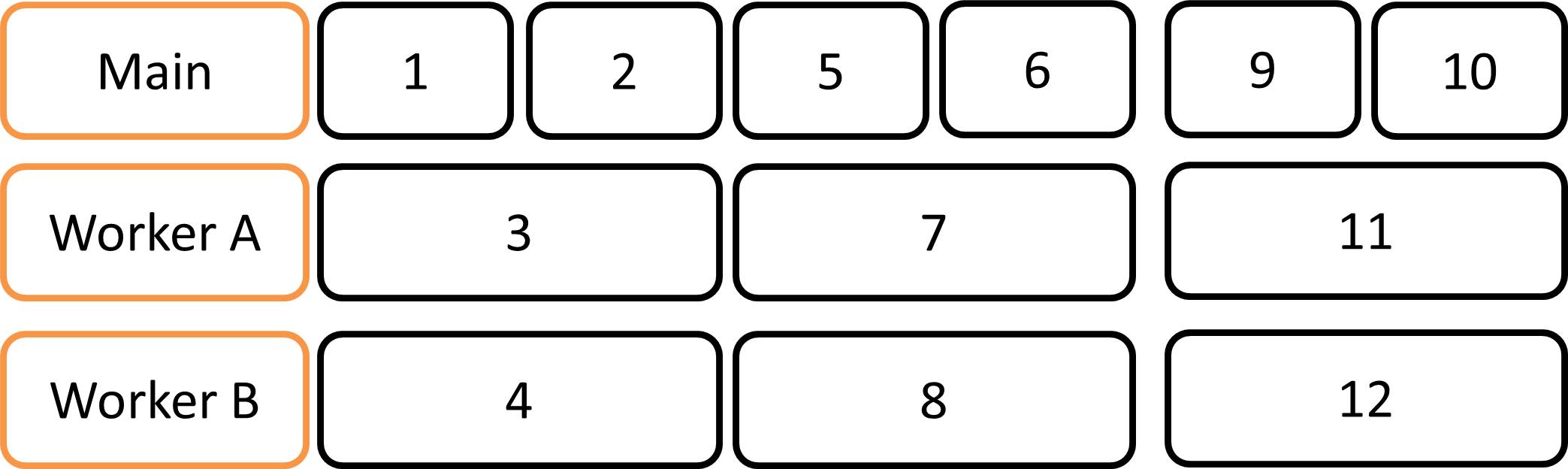}}
\caption{\small ASC with a pair of workers each having 50\% efficiency. The numbers in the boxes are timesteps. The main process computes the first and second timesteps while the workers compute the third and fourth. This lets the main process fast-forward to the fifth timestep and compute both the fifth and sixth timesteps while the workers compute the seventh and eighth. Then the main process can fast-forward to the ninth time-step while the workers computes the eleventh and twelfth, and so on. As a result, the main process finishes the computation twice as quickly as it would natively, achieving 2x speedup with a 50\% computational cache hit rate.}
\label{fig:cpu_efficiency_50}
\end{figure}

Next, let's assume a more realistic scenario where we retain our perfect
prediction, but in which Pin introduces 50\% overhead, meaning that
workers run at half the main process's speed, or alternately, that workers
take twice as long to complete the same amount of work.
Figure~\ref{fig:cpu_efficiency_50} depicts this scenario.

\begin{figure}[!htb]

  \centering
  \begin{minipage}{0.25\textwidth}
    \centering
    \includegraphics[width=\linewidth]{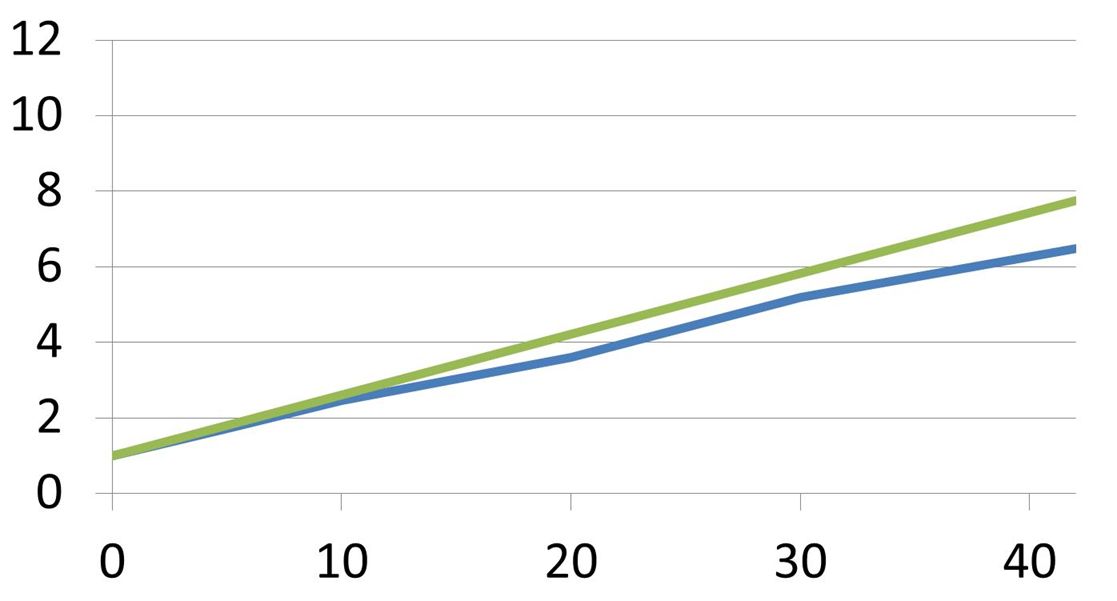}\    \small{readmap}
  \end{minipage}%
  \begin{minipage}{0.25\textwidth}
    \centering
    \includegraphics[width=\linewidth]{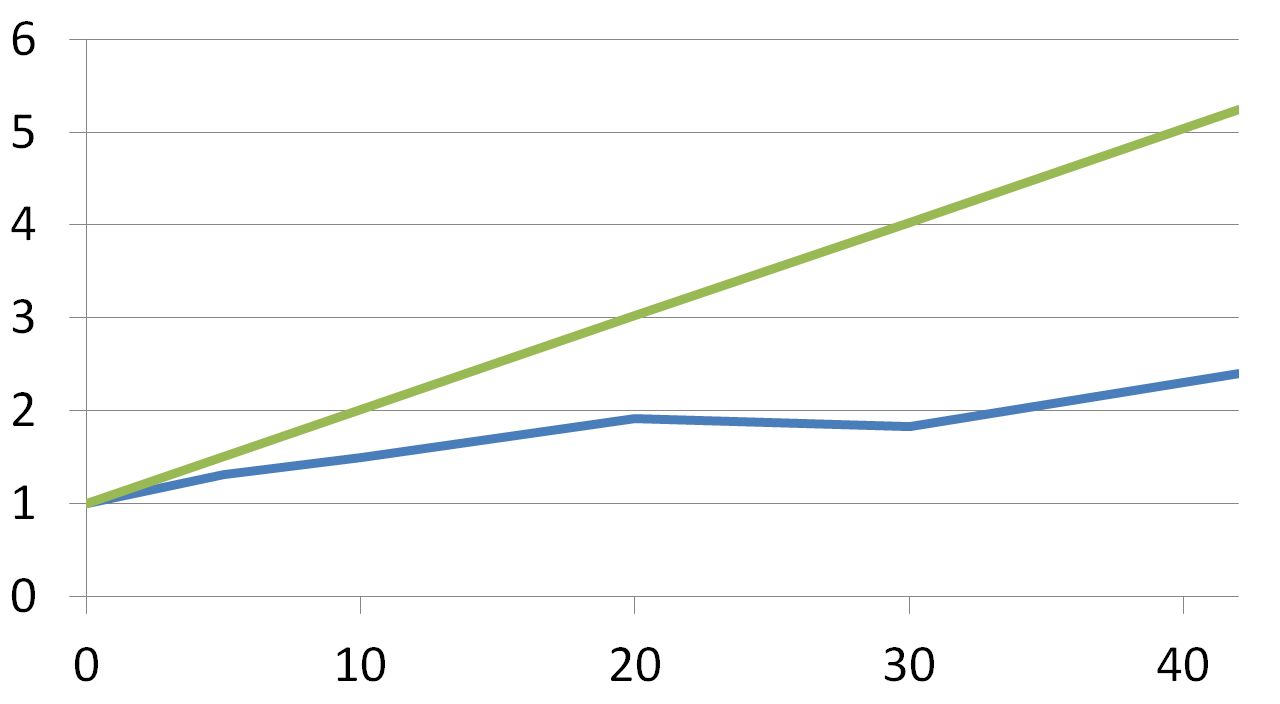}\     \small{gemm}
  \end{minipage}

  \centering
  \begin{minipage}{0.25\textwidth}
    \centering
    \includegraphics[width=\linewidth]{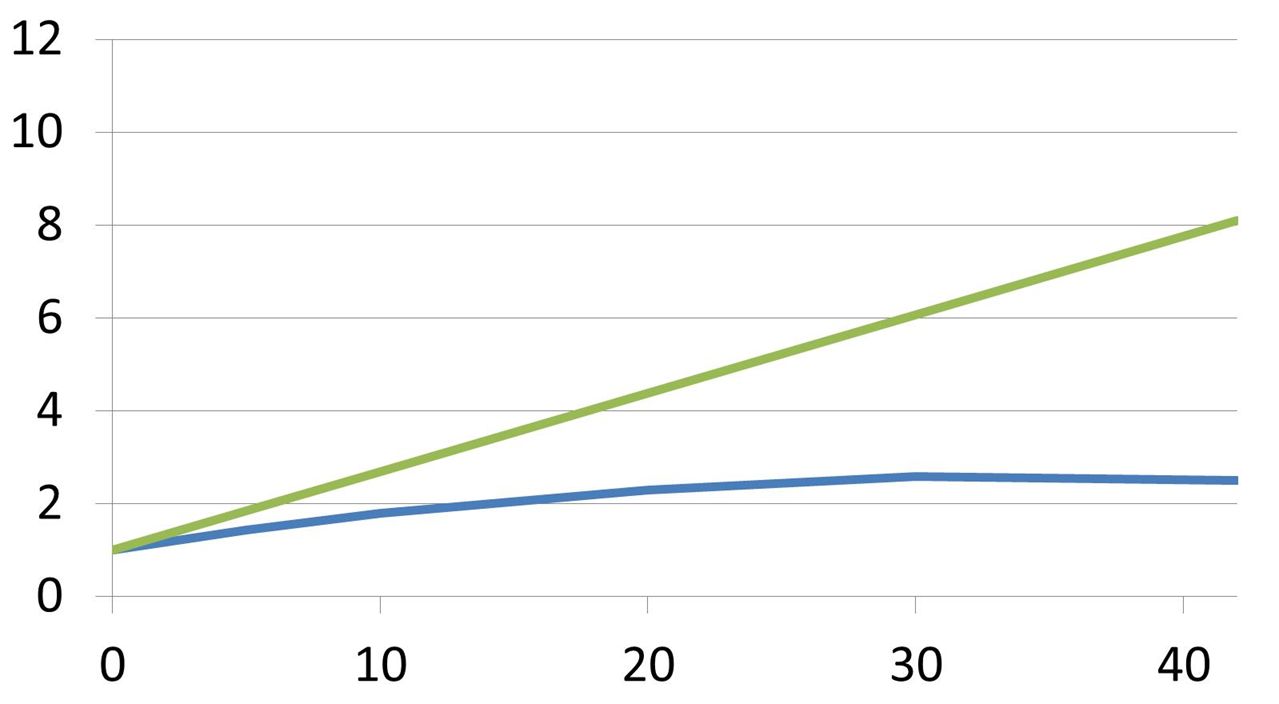}\     \small{cov}
  \end{minipage}%
  \begin{minipage}{0.25\textwidth}
    \centering
    \includegraphics[width=\linewidth]{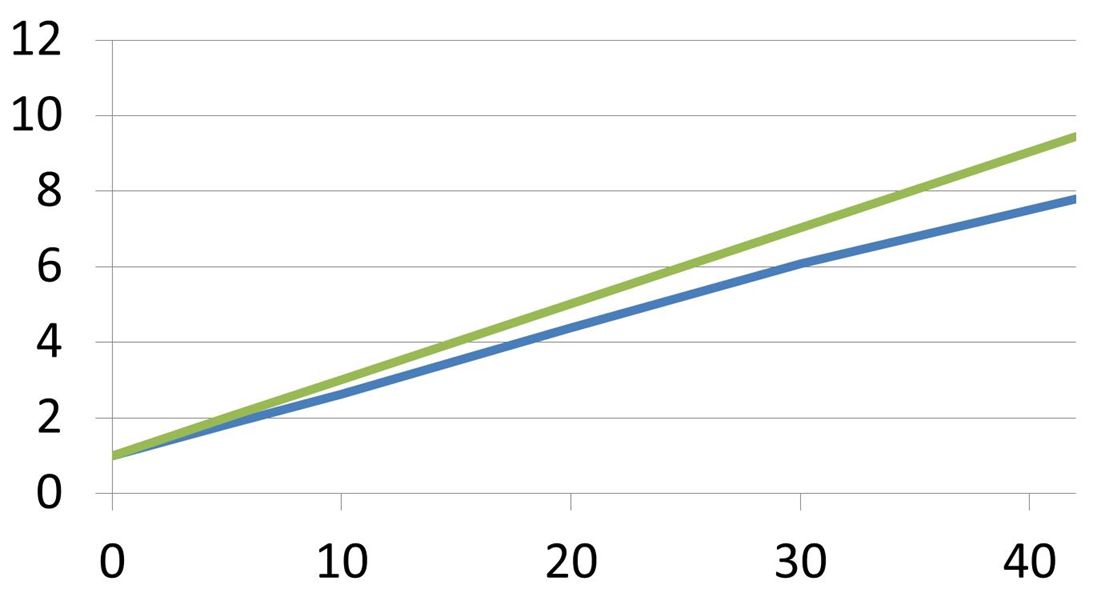}\     \small{3sum}
  \end{minipage}

  \centering
  \begin{minipage}{0.25\textwidth}
    \centering
    \includegraphics[width=\linewidth]{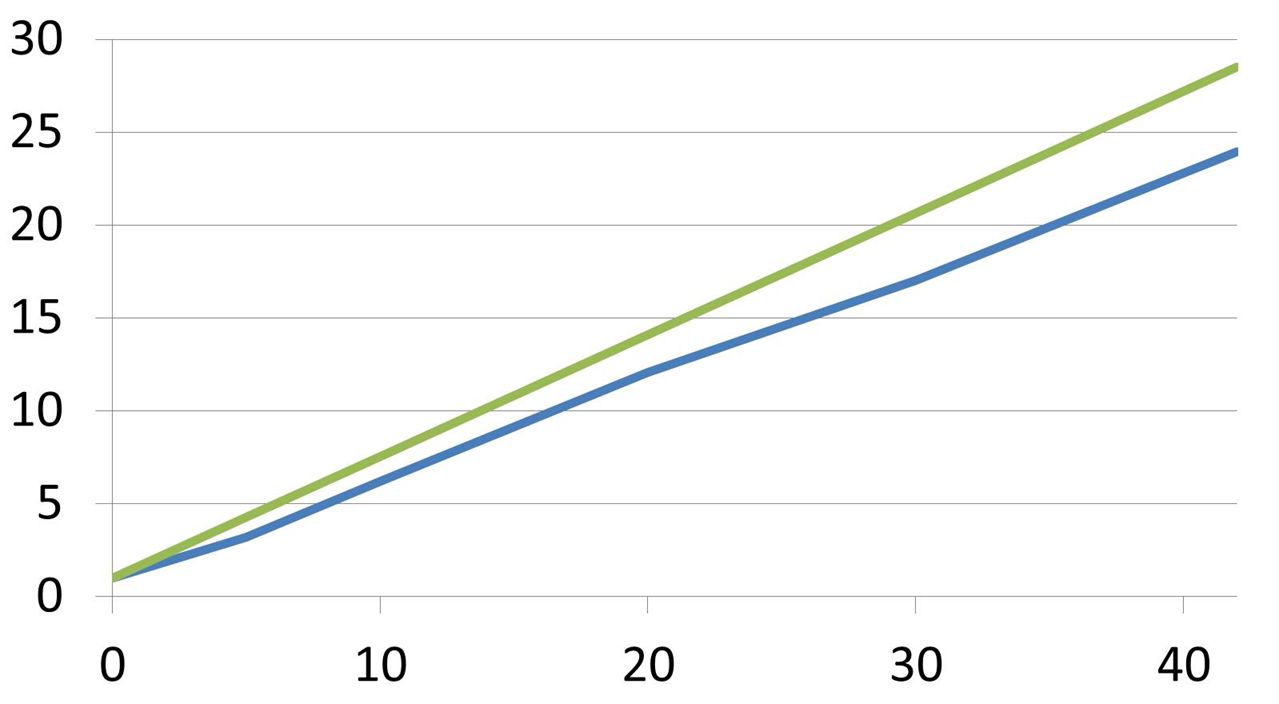}\     \small{ising}
  \end{minipage}%
  \begin{minipage}{0.25\textwidth}
    \centering
    \includegraphics[width=\linewidth]{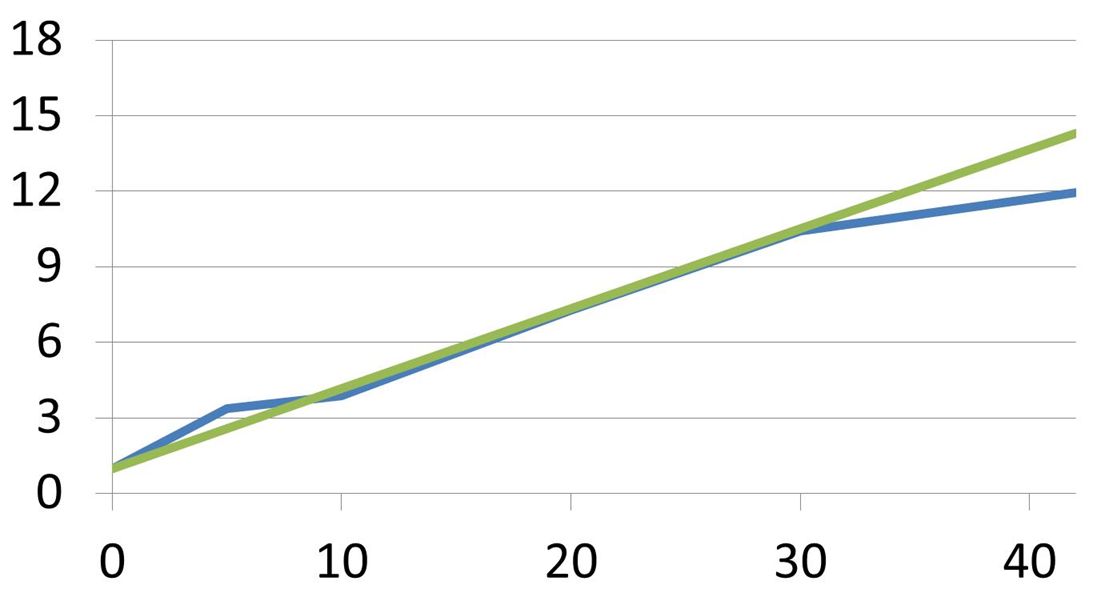}\     \small{collatz}
  \end{minipage}

\caption{\small Diagrams of the actual speedups achieved on all kernels (blue) relative to the maximum speedups possible given the kernels' CPU efficiencies (green).}
    \label{fig:relativekernelanalysis}
\end{figure}

When we had 100\% efficiency, we were able to produce 2x speedup with two cores.
However, when the workers operate at 50\% efficiency, it takes three cores
to achieve the same 2x speedup.
More generally, if a kernel's CPU efficiency is $e$, then the \textit{maximum speedup}
it can attain for $n$ workers is $1 + en$.
We use this formula in Figure~\ref{fig:relativekernelanalysis} to compare
the actual speedups attained by each kernel with the maximum speedups given
their CPU efficiencies.

\begin{table}
\centering
\begin{tabular}{l r}
 \toprule
Kernel & CPU Efficiency\\
\midrule
\texttt{3sum} & 20.1\% \\
\texttt{readmap} & 16.1 \% \\
\texttt{collatz} & 31.7 \% \\
\texttt{ising} & 65.4 \% \\
\texttt{gemm} & 10.1\% \\
\texttt{cov} & 16.9\% \\
 \bottomrule
\end{tabular}
\caption{CPU efficiencies of all kernels. CPU efficiencies are defined as the ratio of the runtime of the kernel when uninstrumented to when instrumented.}\label{table:kernelcpuefficiency}
\end{table}

As we can see from the figures, four of the kernels come close to achieving
their maximum speedup.
The two outliers are the poorly performing kernels, \texttt{cov} and
\texttt{gemm}.
Note that these are also the kernels with extremely large state spaces.
So, what is it about large state spaces that inhibit scaling?

\texttt{gemm} uses well over a gigabyte of state space,
and \texttt{cov} users several hundred megabytes per worker.
The overhead incurred from gathering, scattering, and performing
computational cache operations on these gigantic memory spaces causes
\texttt{cov} and \texttt{gemm} to attain less speedup than they would otherwise.
This is worsened by slowdown from CPU cache contention between all of the
workers, making them compute somewhat more slowly than they would otherwise.
In \texttt{cov} these problems are exacerbated by the nonuniform distance
between breakpoints explained in Section \ref{section:absoluteresults},
which makes optimal scheduling of workers difficult and wastes large amounts
of worker time.
This is the limiting factor for all our large memory benchmarks:
\texttt{trmm, syr2k, 2mm, 3mm}.

Returning to the four kernels that produce speedup
within a small factor of the theoretical maximum given their CPU efficiencies,
two of these kernels have especially interesting results.
\texttt{collatz} actually attains \emph{more} speedup than the
``theoretical maximum'' for small numbers of cores,
because it can reuse speculations left in the cache from earlier predictions.
That is, if every value tested eventually converges to the value 1, then we
know that the last calculation performed for a given value is $2/2$, the
calculation before that is $4/2$, the one before is $8/2$, the one
before that is $16/2$, and the one before that is either $32/2$ or $3*5+1$.
NewAge learns these patterns and is able to fast forward through increasingly
long convergence paths.
This automatic memoization was also present in the original ASC implementation 
~\cite{waterland2014asc}.

\texttt{ising}, meanwhile, attains near-optimal speedup while having extremely
high CPU efficiency, demonstrating that NewAge does not rely on its kernels
having poor CPU efficiencies to scale.
This suggests that improving our CPU efficiency via a lower overhead
data dependency tracking mechanism will not harm NewAge's scalability.
All four kernels' speedups are close to our predicted maximums, given their
CPU efficiencies.
This is, perhaps, the strongest validation of the ASC architecture to date:
if we account for the CPU efficiencies of the workers, our system achieves
speedups comparable to other parallelization systems on kernels with
modest memory overhead.
More importantly, speedups on all four kernels scale near-linearly with an
increasing number of cores.
This suggests that ASC can achieve its goal of providing near-linear
automatic parallelization of programs.

\subsection{Computational Cache Hit Rates} \label{section:absoluteresults}

Lastly, we examine the efficacy of our predictors by examining our
computational cache hit rates.
Computational cache hit rates are defined as the percentage of lookups
(including both synchronous and asynchronous iterated lookups as per
Section \ref{section:newcache}) into NewAge's computational cache (the one described in \ref{section:newcache})  that
hit.
Computational cache hit rates can also be thought of as the percentage
of computation done by the workers instead of on the main process.
A computational cache hit rate of 80\% means that the workers did 80\% of the
overall computation.
Computational cache hit rates can be used to estimate speedup.
For example, a computational cache hit rate of 80\% implies 5x speedup,
because only 20\% of the overall computation is performed by the main process,
while the remaining 80\% is done ``for free'' in parallel by workers on
auxiliary cores.
High levels of overhead can cause speedup to be less than that estimate,
however.

\begin{figure}
\includegraphics[width=\linewidth]{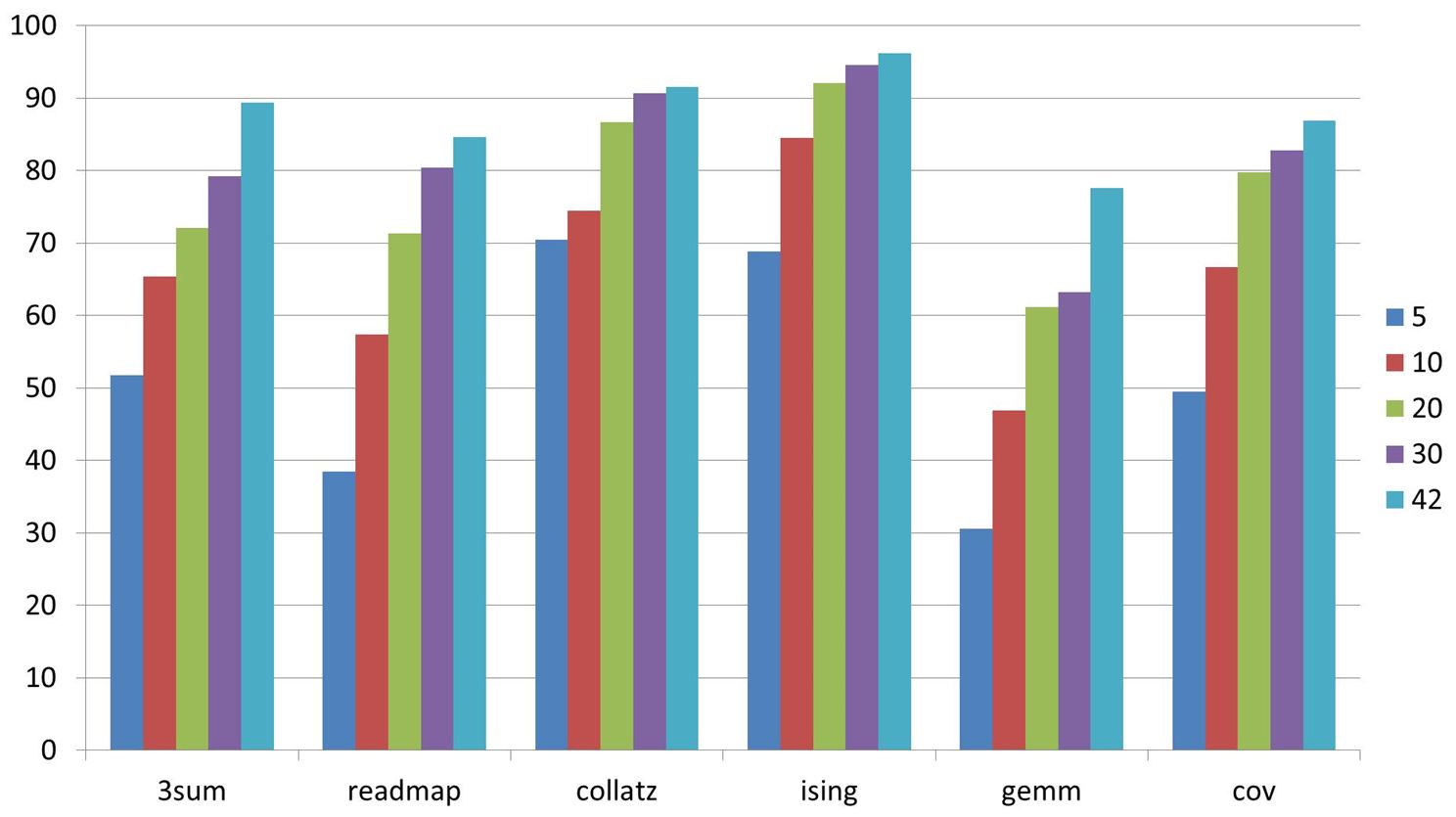}
\caption{\small Computational cache hit rates for all kernels as a function of
the number of workers. Computational cache hit rates are defined as the percentage of lookups into NewAge's computational cache that hit.}
 \label{fig:cache_hit_rates_all}
\end{figure}

Figure \ref{fig:cache_hit_rates_all} presents the computational cache
hit rates for all six kernels as a function of the number of workers.
Four of the kernels---\texttt{readmap}, \texttt{collatz}, \texttt{ising}, and \texttt{gemm}--- have computational cache hit rates exactly as we expect.
That is their hit rate is equal to the percentage of computation done by the
workers instead of by the main process as explained.
For three of these, the computational cache hit rate serves as a good estimator of the speedup we report in Section \ref{section:scalability_results}
(the exception, \texttt{gemm}, achieves less speedup than one would expect due to
complications from its massive memory usage).
Out of these kernels, \texttt{ising} and \texttt{collatz} have the highest
computational cache hit rate, due to their high CPU efficiency and
\texttt{collatz}'s auto-memoization, which allows it to reuse the results
of earlier speculations.
\texttt{readmap} performs slightly worse.
\texttt{gemm} has a lower computational cache hit rate than any of the
preceding three, because of its low CPU efficiency.

For the remaining two kernels, \texttt{3sum} and \texttt{cov}, the
computational cache hit rates are somewhat misleading, because
the distances between breakpoints are non-uniform.
Both programs reach the breakpoint increasingly frequently as they run.
In \texttt{3sum} this occurs because the size of the kernel's search
space shrinks.
In \texttt{cov} this occurs because the height of
the column of the triangular matrix the kernel calculates shrinks.
Because the amount of computation done between breakpoints is
non-uniform, the linear relation between computational cache hit
rate and speedup breaks down.
In \texttt{cov}, the computational
cache hit rate is an overestimate of the amount of speedup achieved,
because later workers add a large number of cache entries that
correspond to little computation but still register as cache hits.
In \texttt{3sum}, which reaches its breakpoint more frequently than
\texttt{cov}, the exact opposite happens, causing the computational
cache hit rate to underestimate the amount of speedup.
Towards the
end of computation, it queries the cache more quickly than the
workers can fill it. This causes a large number of misses, but does
not reduce speedup because those cache misses result in trivial
amounts of computation.

\subsection{Fundamental Limitations of ASC}
\label{section:asclimits}

\begin{figure}
\includegraphics[width=0.85\linewidth]{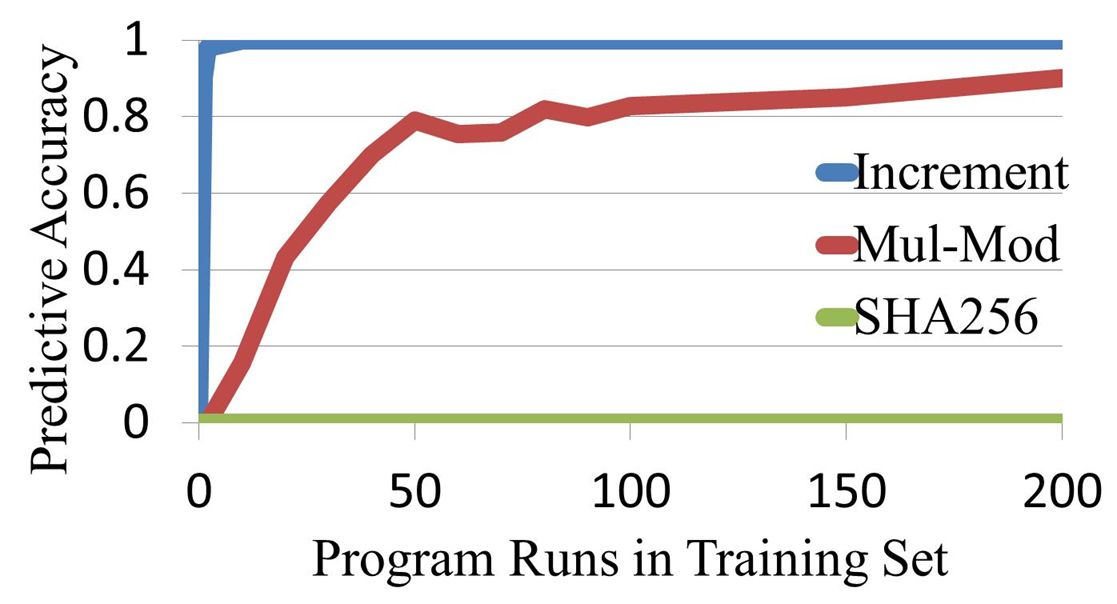}
\caption{\small Predictive accuracy of NewAge's decision tree learner for the program in Listing \ref{source:dependent} as a function of the number of program runs used for training on three different functions $f$.  In \texttt{increment}, $f$ is trivial, returning $j+1$.  In \texttt{mul-mod}, $f$ is more complicated, returning $j$ times a prime number, modulo another prime number.  In \texttt{SHA256}, $f$ is completely unpredictable, returning the SHA256 hash of $j$.}
 \label{fig:asc_limits}
\end{figure}

The primary advantage of ASC with respect to other automatic learning systems is its flexibility.  Other systems mostly use some form of program analysis (both static and dynamic) to analyze dependencies within loops and are limited by the power of the analysis techniques.  ASC, however, converts parallelization into a machine learning problem and is limited only by the ability of its learners to predict future states of a program, such as future iterations of a loop dependent on work done in past iterations.  
This means that programs that cannot be parallelized by a static compiler because of the data dependencies (loop dependent or independent) 
can be parallelized by ASC.
\footnote{Strictly speaking, a compiler can be customized to perform peephole transformations for special cases; but the technique is not generic.}

\lstset{style=customc,basicstyle = \fontencoding{T1}\ttfamily \color{black},caption={\small A simple loop whose iterations are dependent on each other.  ASC can parallelize it if and only if it can predict future values of $f(j)$.  For some functions $f$, this is possible, for others, it is not.}, label={source:dependent}, captionpos=b}

\begin{lstlisting}
void dependmap(int* A, int j, int N) {
  for(int i = 0; i < N; i++) {
    A[i] = j;
    j = f(j);
  }
}
\end{lstlisting}

Listing \ref{source:dependent} presents an example.
Each iteration of the loop depends on previous iterations; specifically on the values of $f(j)$.  While the program seems trivial as written, one can easily imagine a more complex version, where additional operations performed on $j$ before insertion into $A$ make parallelization necessary.  Any form of automatic parallelization dependent on static analysis would have trouble parallelizing Listing \ref{source:dependent}, especially for complicated functions $f$, because of the dependencies between loop iterations.  NewAge, however, can parallelize Listing \ref{source:dependent} for functions $f$ whose future values it can predict.
Figure \ref{fig:asc_limits} illustrates this.
For a simple function $f$, such as $f(j) = j+1$, NewAge achieves 100\% predictive accuracy and fully parallelizes the program within a couple of iterations.  For a more complex function, such as $f(j) = (p_1 * j) \mod p_2$, where $p_1$ and $p_2$ are nontrivial primes, NewAge takes a bit longer to be able to predict future values of $f(j)$, but accuracy eventually reaches 100\%.  But functions whose raison d'etre is to be unpredictable, such as $f(j) = \text{SHA256}(j)$, are impossible for NewAge or any ASC implementation to parallelize; it would be very bad if cryptographic hashes could be predicted efficiently.

\section{Related Work}
\label{section:relatedwork}

As the goal of the ASC approach is automatic parallelization of single-threaded
code, it shares history with, and has drawn ideas
from, a large body of previous work.
We divide this work into two broad categories:
compiler parallelization and binary parallelization;
we note that there are a small number of
speculation-based systems~\cite{Nightingale06}
that do not fit neatly into our categorization.
We further divide each of these categories into their
static and dynamic versions;
ASC is a dynamic, binary parallelization system.
We wrap up this section with a discussion of various
approaches to obtaining the dependency information that
is crucial to ASC's ability to speed up many programs.

\subsection{Compiler Parallelization}

The most traditional method of parallelization is static
compiler parallelization.
This can take different forms.
The most frequently used approach is to write the program in
a parallel language, e.g., Cilk~\cite{blumofe1995cilk}, or using
a parallel runtime, e.g., OpenMP~\cite{dagum1998openmp}.
OpenMP uses pragmas to provide an API to the compiler
to indicate opportunity for parallelization.
Although not entirely automatic, OpenMP does the work of
parallelization, with the programmer only needing to provide hints.
Cilk is a superset of C/C++ that introduces keywords for forking
and joining threads and provides a scheduler to use them
properly.
Again, this is not fully automatic, but makes it relatively
easy for programmers to parallelize otherwise sequential code.

Fully automatic static compilation requires that the compiler recognize
parallelizable loops, detect that the loop computations are independent~\cite{adve95, kennedy01},
and then introduce the proper thread structure to leverage the
opportunity.
For example, the Intel Compilers will automatically
generate OpenMP pragmas for C++ and Fortran, when they
detect loops that can be safely and efficiently executed in parallel~\cite{intelcompiler}.
Such static methods are limited to parallelizing loops where the
compiler is certain of the absence of the data dependencies.
Compilers used in high performance computing employ sophisticated
dependent analysis techniques that include identifying important programming patterns
and transforming them to equivalent programs that are amenable for parallelization~\cite{ibmxl, cray, pgi}.
However, these techniques only handle regular loops and quickly degenerate
in presence of less regular loops or non-standard patterns.

A special case of automatic parallelization is automatic vectorization 
where multiple pairs of scalar operands are operated on at once.
Such compilers use data dependence analysis to check the feasibility of
such transformations and output a vectorized instruction if
possible~\cite{llvm, gcc, intelcompiler}. 
The granularity of parallelization in vectorization is much finer than that
aimed at by ASC and is thus complementary to ASC.
Moreover, ASC benefits from vectorized instructions, exploiting them heavily
in its internal sparse state vector representation.
  
Thread-Level Speculation (TLS) provides a more powerful alternative.
TLS parallelizes code with ambiguous dependencies by making assumptions
about the dependencies, then uses hardware to check those assumptions at runtime.
If the speculations are correct, the computation is used;
if they are not, they can be ignored~\cite{steffan2000scalable, hertzberg2009runtime}.
Hybrid analysis, which is similar to TLS, performs relatively liberal static
analysis then does runtime failsafe checking in software~\cite{rus2003hybrid}.
Both hybrid and TLS techniques are similar to ASC in that they
make use of speculation, but they differ in their a priori use of
static analysis, and in the case of TLS, of hardware support.

There are other hybrid approaches that use both static  and
dynamic analysis.
One approach is to pipeline loops into separate threads
that do not have cyclic dependencies~\cite{dswp,sdswp}, sometimes with
the help of software transactional memory~\cite{raman2010speculative}.
This is unpredictable due to the complexity of the pipelining
transformation, though, and can just as easily slow a program
down as speed it up.
Another technique, and the one most similar to ours, is to have the
compiler parallelize a loop while dynamically transmitting information
about data dependency between cores to guarantee synchronization
and correctness.
The HELIX compiler uses this approach~\cite{campanoni2012helix}.
Like our approach, HELIX can parallelize a variety of programs.
However, HELIX remains dependent on static analysis, and thus
regular data access patterns, and it can easily be crippled by
communication overhead.

\subsection{Binary Parallelization}

The limitations of compiler parallelization have inspired a variety of
alternatives that operate directly on binaries, whether statically or
dynamically.
Like ASC, conventional binary parallelization systems lose access to
the sometimes-valuable semantic information of source code.
In exchange, they also lose their dependency on having access to
the source code and, if they are dynamic, gain access to runtime
information.

The variety of binary parallelization techniques is, if anything,
even greater than the variety of compiler parallelization techniques.
One approach is direct binary translation, which transforms a
sequential binary into a parallelized one~\cite{kotha2010automatic,yardimci}.
This has many of the same advantages and disadvantages as static compiler
parallelization---it is extremely powerful when it works
(even more so than compiler parallelization, as it does not require
the source code and has access to the library binaries), but does not
work on many programs due to the limits of static analysis~\cite{kotha2010automatic}.
A closely related technique is \textit{slicing}, which uses static
analysis of data flow to divide a program into parallel slices, while
using speculation to work around rare dependencies~\cite{wang2009dynamic}.

Alternatively, Dynamic Binary Parallelization~(DBP) techniques abandon
static analysis altogether.
One of the first examples of this work is Dynamo~\cite{bala2000dynamo}.
Although Dynamo was not a parallelization scheme per se,
rather optimizer that identified critical sections in code as
it ran and made them faster, it inspired work in parallelization.
For example, DBP monitors program execution for frequently-executed
\textit{hot traces} that can be parallelized as they reappear~\cite{yang2011feasibility}.
Such approaches share ASC's
predict-speculate-fast-forward paradigm, but are limited by their
reliance on hot traces alone, unlike ASC's more general-purpose prediction.

\subsection{Dynamic Instrumentation}

Our ASC implementation relies on dynamic instrumentation to identify
the bits on which a computation depends.
This is an area with a rich research history.
One approach is full transformation, where the code is translated to
an intermediate representation before being run on a virtual machine or simulator.
Valgrind~\cite{nethercote2007valgrind} is the most prominent example of
this approach.
Using a virtual machine, as the initial ASC prototype did~\cite{waterland2014asc},
makes instrumentation much easier, as the virtual
machine can simply record whatever it is simulating.
However, the tradeoff for this ease of instrumentation is
poor performance.
Valgrind is, in the best case, four or five times slower than native
code and is much worse with extensive instrumentation, while the
first ASC prototype was multiple orders of magnitude slower than native
execution.

Another approach to instrumentation, and the one we adopt, is to
use an extremely lightweight just-in-time compiler (JIT) to run the
binary code and insert instrumentation as appropriate.
DynamoRIO~\cite{bruening2003infrastructure} (based off of Dynamo)
and Pin~\cite{luk2005Pin} are prominent examples of this approach.
Both tools have extremely low base overhead (on the order of 10\% for
DynamoRIO and 30\% for Pin),
due to JIT optimization, and both allow the insertion
of a wide range of instrumentation and analysis code.
We chose Pin for its more powerful API and
instrumentation capabilities.

\section{Conclusions and Future Work}
\label{section:conclusion}

\subsection{Summary}

We present a new implementation of the ASC architecture, a powerful approach for automatic parallelization.  We demonstrate that ASC is capable of working on unmodified binaries without the use of static analysis, providing more potential power than other automatic parallelization systems.  We demonstrate ASC's ability to automatically parallelize a variety of kernels.  These kernels include fundamental parallelizable computations, such as maps or matrix multiplication, along with problems other automatic parallelization systems struggle with, such as linked list navigation and auto-memoization.  Moreover, we show that ASC's speedups scale linearly with the number of cores, allowing it to work efficiently on systems with large numbers of cores.  This makes ASC potentially useful for many scientific computing applications, where researchers not trained in parallelization techniques need to run extremely computationally intensive programs on extremely powerful computers.

\subsection{Future Work}

We were pleasantly surprised at how well the ASC approach works in practice.
These encouraging results suggest opportunities for even more
powerful future implementations.

The most promising avenue of future work is improving on the
computational efficiency of dependency tracking; Pin is effective, but
expensive.
There are at least two potential directions to explore: specialized hardware,
perhaps similar to transactional memory hardware, that records information that
we collect in the read and write masks, and virtual memory integration to more easily
obtain information on what areas of memory were read from and written to.

While dependency tracking dramatically reduces the size of the state space for some kernels, it does not minimize it.  As a result, there are many programs that are parallelizable but that ASC cannot speed-up.  For example, ASC cannot speed up an accumulator that sums the elements in an array, because that requires predicting the values of the sum.  If ASC's analysis revealed that,
for example, the value of the sum was never queried but only added to itself (an associative and commutative operation), ASC could replace the prediction of sum with an addition of partial values.

Section \ref{section:scalability_results} showed that NewAge's performance relative to theoretical maximums is good, but we believe it should be possible to do better, particularly on large footprint programs. Better code optimization for operations such as scatter, gather, and predict should reduce the constant factor of our linear time-dependence on the memory size of worker processes.  This should dramatically improve performance on high-memory kernels such as \texttt{gemm}.

The completion of any of these will make ASC far more powerful. The development of all of them---which is completely possible---will make ASC a fully mature technology capable of outperforming and outscaling other automatic parallelization approaches.

\bibliographystyle{acm}
\bibliography{thesis_bib}

\end{document}